\documentclass{PoS}

\usepackage{amsmath, multirow, tikz, graphicx, mwe}

\usepackage[natbib=true, sorting=none, maxcitenames=3, maxbibnames=3, style=numeric, backend=bibtex8, doi=false,isbn=false]{biblatex}
\DeclareFieldFormat*{title}{\emph{#1}}

\title{Multiwavelength observations of the blazar BL Lacertae: a new fast TeV $\gamma$-ray flare}

\ShortTitle{Multiwavelength observations of the blazar BL Lacertae: a new fast TeV $\gamma$-ray flare}

\newcommand\apjs{ApS}%
\newcommand\apjl{ApJL}%
\newcommand\aap{A\&A}%
\graphicspath{{../}}

\author{\speaker{Q.~Feng}$^a$ for the VERITAS Collaboration\thanks{veritas.sao.arizona.edu}, S.~G.~Jorstad$^{b,c}$, A.~P.~Marscher$^b$, M.~L.~Lister$^d$, 
            Y.~Y.~Kovalev$^{e,f}$, A.~B.~Pushkarev$^{g,e}$, T.~Savolainen$^{h,i,f}$, I.~Agudo$^i$, S.~N.~Molina$^j$,  J.~L.~Gomez$^j$,  V.~M.~Larionov$^c$,  G.~A.~Borman$^g$,  
            A.~A.~Mokrushina$^c$, and P.~S.~Smith$^k$ \\
         \llap{$^a$}Physics Department, McGill University, Montreal, QC H3A 2T8, Canada\\
         E-mail: \email{qi.feng2@mcgill.ca} \\
\llap{$^b$}Institute for Astrophysical Research, Boston University, 725 Commonwealth Avenue, Boston, MA 02215, USA \\
\llap{$^c$}Astronomical Institute, St.Petersburg State University, Universitetskij Pr. 28, Petrodvorets, 198504 St.Petersburg, Russia\\
\llap{$^d$}Purdue University, 525 Northwestern Avenue, West Lafayette, IN 47907, USA \\
\llap{$^e$}Astro Space Center of Lebedev Physical Institute, Profsoyuznaya 84/32, 117997 Moscow, Russia \\
\llap{$^f$}Max-Planck-Institut f\"ur Radioastronomie, Auf dem H\"ugel 69, 53121 Bonn, Germany \\
\llap{$^g$}Crimean Astrophysical Observatory, 98409 Nauchny, Crimea, Russia \\
\llap{$^h$}Aalto University Mets\"ahovi Radio Observatory, Mets\"ahovintie 114, FI-02540 Kylm\"al\"a, Finland \\
\llap{$^i$}Aalto University Department of Electronics and Nanoengineering, PL 15500, FI-00076 Aalto, Finland \\ 
\llap{$^j$}Instituto de Astrof\'{\i}sica de Andaluc\'{\i}a (CSIC), Apartado 3004, E--18080 Granada, Spain \\
\llap{$^k$}Steward Observatory, University of Arizona, Tucson, AZ 85716, USA\\
}

\abstract{
Observations of fast TeV $\gamma$-ray flares from blazars reveal the extreme compactness of emitting regions in blazar jets. Combined with very-long-baseline radio interferometry measurements, they probe the structure and emission mechanism of the jet. 
We report on a fast TeV $\gamma$-ray flare from BL Lacertae observed by VERITAS, with a rise time of about 2.3 hours and a decay time of about 36 minutes. %It was the second from this lower-frequency peaked BL Lac object. 
The peak flux at $>$200 GeV measured with the 4-minute binned light curve is $(4.2 \pm 0.6) \times 10^{-6} \;\text{photons} \;\text{m}^{-2}\; \text{s}^{-1}$, or $\sim$180\% the Crab Nebula flux. 
Variability in GeV $\gamma$-ray, X-ray, and optical flux, as well as in optical and radio polarization was observed around the time of the TeV $\gamma$-ray flare. A possible superluminal knot was identified in the VLBA observations at 43 GHz. 
The flare constrains the size of the emitting region, and is consistent with several theoretical models with stationary shocks. 
}
\FullConference{35th International Cosmic Ray Conference - ICRC2017\\
		12-20 July, 2017\\
		Bexco, Busan, Korea}

\begin{document}
%%%%%%%%%
% Intro
%%%%%%%%%
  \setlength{\abovecaptionskip}{0pt}

\vspace*{-1.2cm}
\section{Introduction} \label{sec:intro}
\vspace{-0.2cm}

%BL Lac objects belong to a subclass of radio loud active galactic nuclei (AGN), known as blazars. They are characterized by featureless optical spectra, non-thermal broadband spectra, and rapid variability, which jointly suggest that their emission originates in relativistic jets closely aligned to our line of sight \cite[e.g.][]{Schlickeiser96}. 

Some TeV blazars exhibit fast $\gamma$-ray variability, the timescale of which can be as short as a few minutes at very high energies (100~GeV $\lesssim E_{\gamma} \lesssim$ 100~TeV; VHE). Such variability has been observed in several BL Lac objects, %\cite[e.g.][]{Gaidos96, Aharonian07, Albert07, Aleksic11}, 
including the prototype BL Lacertae (BL Lac hereafter) \cite{Arlen13}, located at a redshift of $z=0.069$ (an angular scale of $\approx1.3$~pc/mas). %\cite{Miller77}. %EG2015IC310 
%
%The observed rapid gamma-ray variability of TeV blazars implies very compact emitting regions, as well as low gamma-ray attenuation caused by pair production on infrared/optical photons.

Long-term monitoring of BL~Lac %by multiple TeV gamma-ray instruments 
indicates that the source is not detectable in the TeV $\gamma$-ray band except during flaring episodes, %with its flux reaching $>$100\% of the Crab Nebula flux (C.~U.) above 1 TeV in 1998 \cite{Neshpor01}, $\sim$3\% C.~U. above 200~GeV in 2005 \cite{Albert07BLLac}, and 
the most recent of which exhibited a flux of $\sim$125\% of the Crab Nebula flux (C.~U.) above 200~GeV with a fast variability timescale of $13\pm4$ minutes in 2011 \cite{Arlen13}. 

BL Lac exhibits both stationary radio cores/knots and superluminal radio knots \cite[e.g.][and references therein]{Hervet16}. % \cite{Lister13}. 
Possible associations between the variability of superluminal radio knots and $\gamma$-ray flares have been investigated for the source \cite[][]{Marscher08, Arlen13}. %, and other blazars \cite[e.g.][]{Rani14, Max-Moerbeck14}. 

On 2016 Oct 5, the Very Energetic Radiation Imaging Telescope Array System (VERITAS) observed a sub-hour TeV $\gamma$-ray flare from BL Lac.  
A series of observations with the Very Long Baseline Array (VLBA) were taken at 43~GHz and 15.4~GHz %(and 86~GHz?) 
over the span of a few months, revealing that a possible knot structure emerged around the time of the TeV $\gamma$-ray flare. %(see Subsection~\ref{sec:VLBA}).
In this work, we report on the results of the aforementioned and other MWL observations, and discuss their implications. 
%The cosmological parameters assumed through out this paper are $\Omega_m=0.27$, $\Omega_{\Lambda}=0.73$, and $H_0=70\;\text{km}\;\text{s}^{-1}\;\text{Mpc}^{-1}$. %\cite{Larson11}. 
%At the redshift of BL Lac, %the luminosity distance and the angular size distance are about $311$~Mpc and $273$~Mpc, respectively, and 
%the angular scale is about $1.3$~pc/mas. 

%%%%%%%%%
% Observations and data analysis
%%%%%%%%%
\vspace{-0.3cm}
\section{Observations, Data Analysis, and Results}
\label{sec:obs_data}
\vspace{-0.2cm}
%%%%%%%%%
\subsection{VERITAS}
\label{sec:VERITAS}
%\subsection{VHE gamma-ray observations with VERITAS}
%%%%%%%%%
%
\vspace{-0.05cm}
VERITAS is an array of four imaging atmospheric Cherenkov telescopes located in southern Arizona \cite[see e.g.][for details]{Holder11}. %It is sensitive to gamma rays in the energy range from 85 GeV to $>$30 TeV with an energy resolution of $\sim$15\% (at 1~TeV), and capable of making a detection at a statistical significance of 5 standard deviations (5 $\sigma$) of a point source of 1\% of the Crab Nebula flux in $\sim$25~hours. 
%
%Each of the four telescopes is equipped with a 12-m diameter Davies-Cotton reflector comprising 355 identical mirror facets, 
%and a 499-pixel photomultiplier tube (PMT) camera covering a field of view of 3.5$^\circ$ at an angular resolution (68\% containment) of $\sim$0.1$^{\circ}$ (at 1~TeV). Coincident Cherenkov signals from at least two out of the four telescopes are required to trigger an array-wide read-out of the PMT signals. 
%
BL Lac was observed at an elevated TeV $\gamma$-ray flux by VERITAS on 2016 Oct 5, %as part of the long-term-plan snapshot program \cite{Benbow16}, and instantaneously followed up %by VERITAS 
%based on a real-time analysis. The total exposure of these observations amounts to 
with an exposure of 153.5 minutes after data quality selection. %, with zenith angles ranging between 11$^\circ$ and 30$^\circ$. 
The data were analyzed using two independent analysis packages %\cite{Cogan08, Daniel08} 
and pre-determined cuts optimized for lower-energy showers. %\cite[see e.g.][]{Archambault14}. 
A detection with a statistical significance of $70.7\sigma$ was made from the data of that night, with a time-averaged integral flux above 200~GeV of $(2.24 \pm 0.06) \times 10^{-6} \;\text{photons} \;\text{m}^{-2}\; \text{s}^{-1}$.

%%%%%%%%%
%\subsubsection{VHE $\gamma$-ray flux variability and the modelling of the flare profile}
%\label{subsubsec:VHELC}
%%%%%%%%%
\iffalse
\begin{figure}[t]
 \centering 
%\plotone{BLLac_20161005_LC200GeV_VERITAS_MCMC2models.pdf}
%\hspace{-1cm}
%\fig{MC_sim99_no_baseline_moresims_a3p5_v3.pdf}{0.58\textwidth}{}
 \includegraphics[width=0.6\textwidth]{MC_sim99_no_baseline_moresims_a3p5_v3.pdf}
\caption{The VERITAS TeV $\gamma$-ray light curves of BL Lac $>200$ GeV on 2016 Oct 5. The blue dots show the light curve in 4-minute bins, and the red squares show the light curve in 30-minute bins. The green dashed line and shaded region show the best-fit model and its 99\% confidence interval, respectively, using Markov chain Monte Carlo sampling. 
\label{fig:vlc}}
\end{figure}
\fi
%
Figure~\ref{fig:vlc} shows the VERITAS light curves of BL Lac on 2016 Oct 5 with 4-minute and 30-minute bins. %A constant fit to the 4-minute binned light curve yields a p-value of $1.1 \times 10^{-16}$ ($\chi^2$=170.8 for 45 degrees of freedom; DOF), rejecting the hypothesis of steady flux. 
%A gradual rise of the TeV flux by a factor of $\sim$2 followed by a faster decay was observed. 
The measured peak flux of the 30-minute binned light curve is $(3.0 \pm 0.2) \times 10^{-6} \;\text{photons} \;\text{m}^{-2}\; \text{s}^{-1}$, corresponding to $\sim$125\% C.~U., and that of the 4-minute binned light curve is $(4.2 \pm 0.6) \times 10^{-6} \;\text{photons} \;\text{m}^{-2}\; \text{s}^{-1}$, or $\sim$180\% C.~U.
\begin{figure}[t]
\vspace{-0.85cm}
 \centering 
    \begin{minipage}{0.48\textwidth}
        \centering
%        \begin{tikzpicture}
%\node (img) { \includegraphics[width=1.08\textwidth]{MC_sim99_no_baseline_moresims_a3p5_v3.pdf} };
\includegraphics[width=1.07\textwidth]{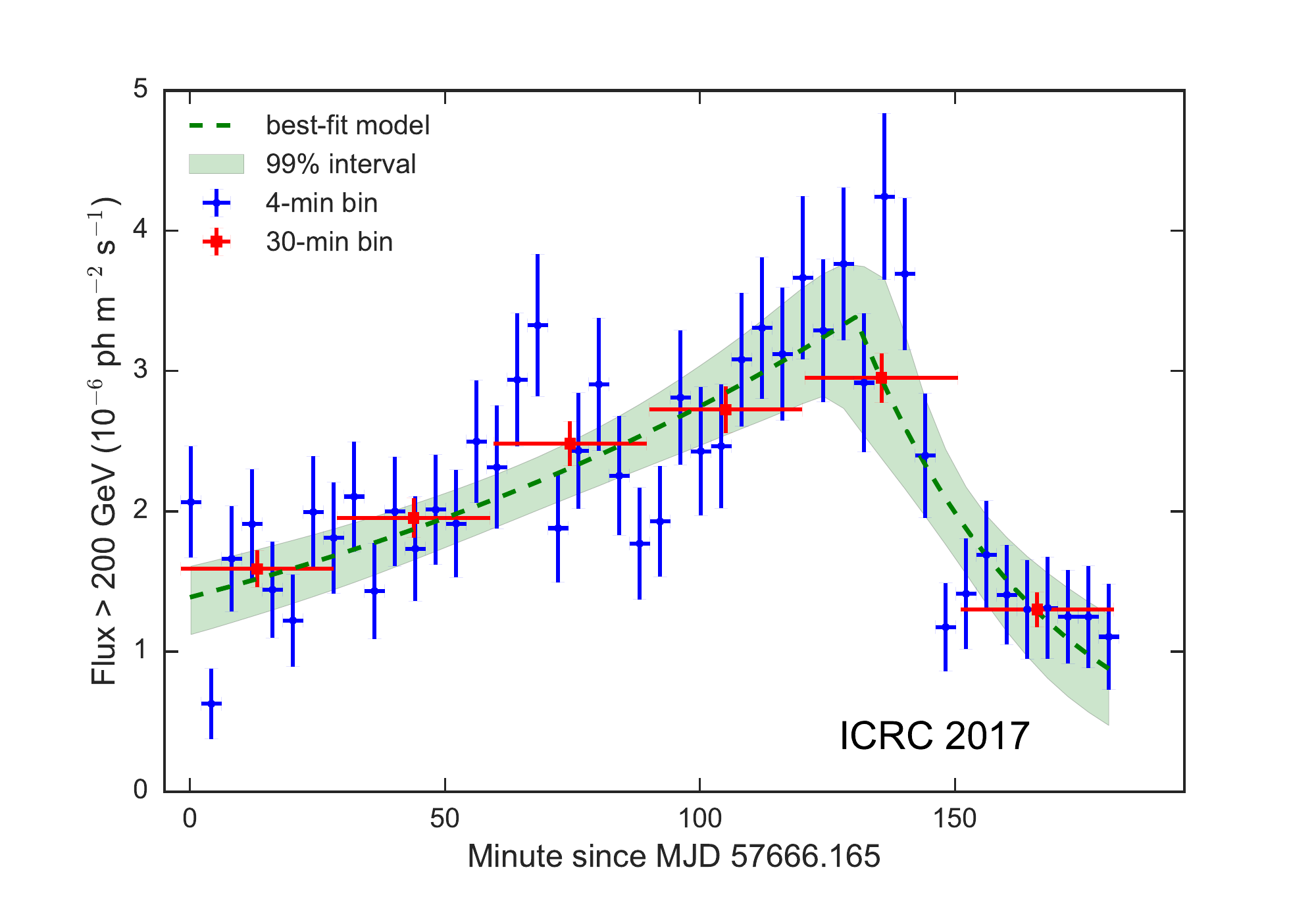}
%\node [below left,text width=3cm,align=center, rotate=45, opacity=0.5] at ([shift={(45:2)}]img){{\Large \bf 2017 ICRC}};
%\node [text width=3cm, align=center, rotate=45, opacity=0.5] at ([shift={(45:0)}]img) {{\Large \bf 2017 ICRC}};
%\node [text width=3cm, align=center, rotate=45, opacity=0.5, color=orange] {{\Large \bf 2017 ICRC}};
%\end{tikzpicture}
 \vspace{-0.85cm}
\caption{The VERITAS TeV {$\gamma$}-ray light curves of BL Lac $>200$ GeV on 2016 Oct 5. The blue dots show the light curve in 4-minute bins, and the red squares show the light curve in 30-minute bins. The green dashed line and shaded region show the best-fit model and its 99\% confidence interval, respectively, using Markov chain Monte Carlo sampling. 
\label{fig:vlc}}
    \end{minipage}\hfill
    \begin{minipage}{0.48\textwidth}
        \centering
%        \begin{tikzpicture}
%\node (img) {   \includegraphics[width=1.0\textwidth]{BLLac_Xray_gamma_SED_2016flare_logpara_deabs.pdf} };
\includegraphics[width=1.02\textwidth]{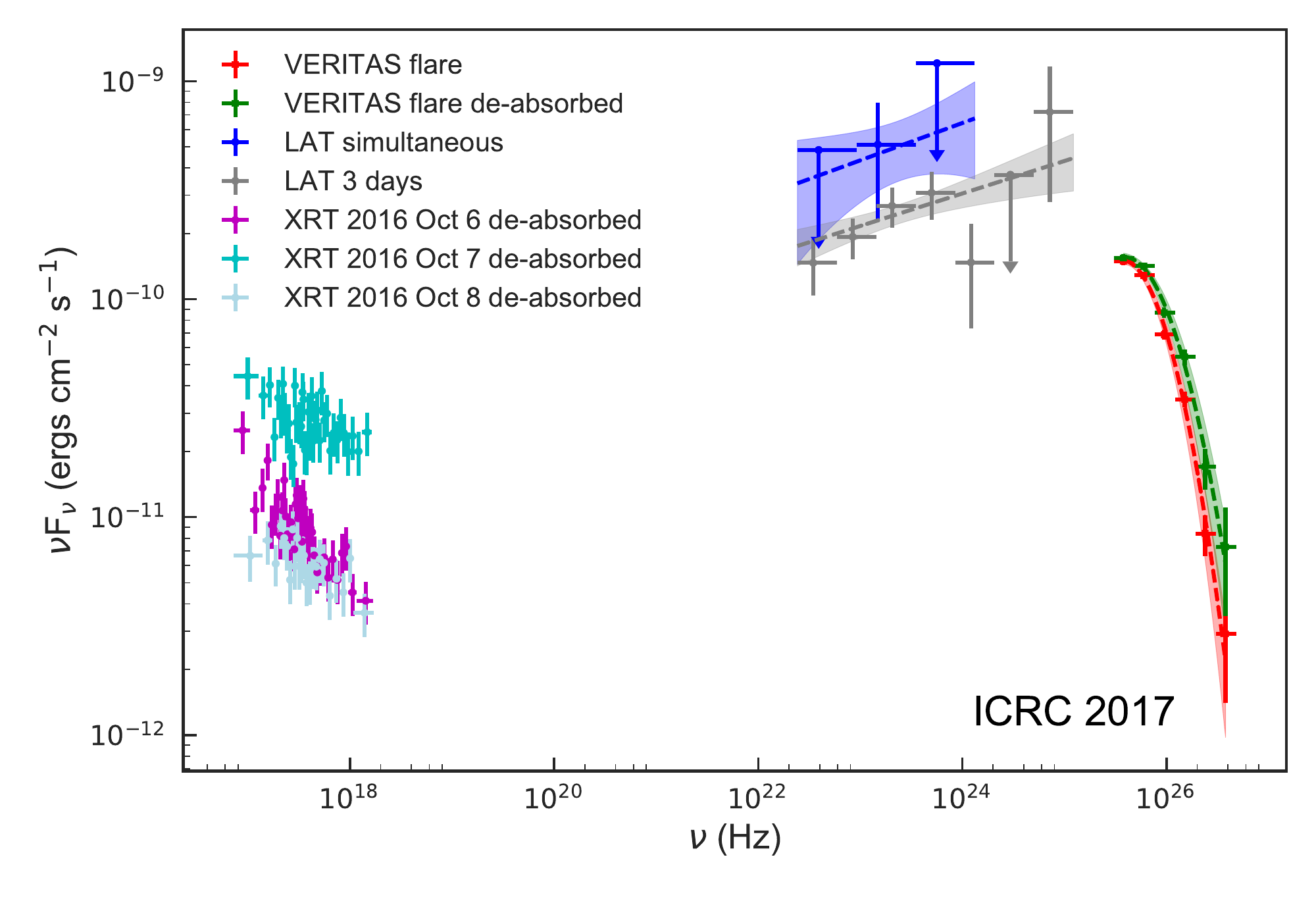}
%\node [text width=3cm, align=center, rotate=45, opacity=0.5, color=orange] {{\Large \bf 2017 ICRC}};
%\end{tikzpicture}
  \vspace{-0.85cm}
 \caption{The $\gamma$-ray and X-ray SEDs measured by {\it Fermi}-LAT, VERITAS, and {\it Swift}-XRT. The {\it Fermi}-LAT SEDs, strictly simultaneous with VERITAS observations on 2016 Oct 5 and from the three days around it, are shown in blue and grey, respectively. %The observed and de-absorbed VERITAS SED averaged over all observations on the flaring night is shown in red and green, respectively. 
The shaded regions represent 1-$\sigma$ confidence intervals derived from the best-fit spectral model. 
%The X-ray SEDs de-absorbed with the best-fit neutral hydrogen column density values on 2016 Oct 6, 7, and 8 are shown in magenta, cyan, and light blue, respectively. 
 %power-law model from the LAT unbinned likelihood analysis. 
 \label{fig:gammaSED}}
    \end{minipage}
    \vspace{-0.6cm}
\end{figure}
%

%We first fitted the VERITAS light curve with 4-minute bins with a constant-flux model, rejecting the constant-flux hypothesis with a p-value of $1.1\times10^{-16}$. %obtaining a reduced $\chi^2$ value of 3.8, corresponding to a p-value of $1.1\times10^{-16}$ and rejecting the constant-flux hypothesis. 
To quantify the rise and decay times of the TeV flare, we fitted the VHE $\gamma$-ray light curve with a piece-wise exponential function %as follows: 
%
%\begin{equation}
%\label{flare_func}
$
F(t) = F_0 e^{-\left|t - t_{\text{peak}}\right|/t_{\text{var}}},
%\begin{cases}
%%%F_{\text{peak}} e^{(t - t_{\text{peak}})/t_{\text{rise}}}, & t \leqslant t_{\text{peak}};  \\
%%%F_{\text{peak}} e^{-(t- t_{\text{peak}})/t_{\text{decay}}}, & t \geqslant t_{\text{peak}}; 
%F_0 e^{(t - t_{\text{peak}})/t_{\text{rise}}}, & t \leqslant t_{\text{peak}};  \\
%F_0 e^{-(t- t_{\text{peak}})/t_{\text{decay}}}, & t \geqslant t_{\text{peak}}; 
%\end{cases} 
$
%\end{equation}
%
where %$F_{\text{peak}}$ 
$F_0$ is the peak flux, $t_{\text{peak}}$ is the peak time, and $t_{\text{var}}$ is the rise or decay time. % $t_{\text{rise}}$ and $t_{\text{decay}}$ are the rise and decay timescales, respectively, on which the flux varies by a factor of $e$. 
%Flare profile \cite{Sato08}: 

The best parameters and their 99\% confidence intervals were determined %from the posterior distributions obtained from 
using Markov chain Monte Carlo (MCMC) simulations, %using the Python package \texttt{emcee}~\cite{Foreman-Mackey13}, 
and are shown in Figure~\ref{fig:vlc}. 
The rise and decay timescales of the flare are determined to be $140^{+25}_{-11}$ minutes and $36^{+8}_{-7}$ minutes, respectively. 

Further VERITAS observations of BL Lac were taken on Oct 6 and from Oct 22 to Nov 19 with 37.6-minute and 294.6-minute live exposure after data quality selection, respectively, neither of which led to a detection of the source. %(at $2.6\sigma$ and $0.9\sigma$, respectively). 
The integral flux upper limits $>$200 GeV at 99\% confidence level on Oct 6 and between Oct 22 and Nov 19 were $2.0\times10^{-7}$  %\;\text{photons} \;\text{m}^{-2}\; \text{s}^{-1}$ 
and $2.8\times10^{-8}  \;\text{photons} \;\text{m}^{-2}\; \text{s}^{-1}$, respectively, assuming a power-law spectrum with a photon index of $3.3$.  

Motivated by the existence of multiple radio emission zones identified in VLBA data (see~Subsection~\ref{sec:VLBA}), %and several multi-zone models for BL~Lac that are consistent with past observations \cite[e.g.][]{Raiteri13, Hervet16}, 
we also fitted the light curve with a model including a constant flux baseline. 
In a multi-zone model, %different zones can be of different sizes and vary independently on different timescales. Therefore, 
it is possible to have a larger emitting zone that varies slowly and can be adequately described by a constant baseline on the timescale considered, and a smaller, more energetic zone that is responsible for the fast flare described by the exponential components. 
With the more complex model, the best decay time is only $2.6^{+6.7}_{-0.8}$ minutes, with a baseline flux of $1.2^{+0.1}_{-0.2}\times10^{-6} \;\text{ph}\; \text{m}^{-2} \text{s}^{-1}$. The baseline flux is higher than the upper limit obtained from the observations taken the following day, indicating the potential slower component varies on the timescale of $\sim$1 day, consistent with the GeV observations. %Although 
We note that with our limited statistics it is not possible to unambiguously reject either model. %based on the statistics. 

%However, regardless of the models and fitting procedures used, the decay timescale of the VHE flare is smaller than an hour. 

%%%%%%%%%
%\subsubsection{The VHE spectrum}
%%%%%%%%%
A power-law fit to the VERITAS spectrum of BL Lac yields a $\chi^2_\text{dof}$ value of 34 and a best-fit photon index of $3.28\pm 0.04$, insufficient to describe the data.  
\iffalse
parameters as follows: 
\begin{equation}
\label{eqVspecPL}
\begin{split}
\frac{dN}{dE}=& (1.03\pm 0.06)\times 10^{-7} %\\ & 
\times \left( \frac{E}{1\text{TeV}} \right)^{(-3.28 \pm 0.04)} %\\ & 
\text{m}^{-2} \text{s}^{-1} \text{TeV}^{-1}.
\end{split}
\end{equation}
\fi
%
A log parabola model with a fixed pivot energy of 0.2 TeV fits the VERITAS spectrum better: 
%\begin{equation}
%\label{eqVspecLP}
$
%\begin{split}
%\frac{dN}{dE}&= (2.22\pm 0.07)\times 10^{-5} %\\ & 
\frac{dN}{dE}= (2.22\pm 0.07)\times 10^{-5} %\\ & 
\times \left( \frac{E}{0.2\text{TeV}} \right)^{\left[ -(2.4 \pm 0.1) -  ( 1.8 \pm 0.3 ) \log_{10} (\frac{E}{0.2\text{TeV}})  \right]} %\\ & 
\text{m}^{-2} \text{s}^{-1} \text{TeV}^{-1}, 
%\end{split}
$
%\end{equation}
with a $\chi^2_\text{dof}$ value of 1.6. 
%
%After de-absorbing the VHE spectrum using the optical depths in \cite{Dominguez11}, the best-fit log parabola model becomes: 
%\begin{equation}
%\label{eqVspecLPdeabs}
%\begin{split}
%$
%\frac{dN}{dE}&= (2.36\pm 0.07)\times 10^{-5} %\\ & 
%\frac{dN}{dE}= (2.36\pm 0.07)\times 10^{-5} %\\ & 
%\times \left( \frac{E}{0.2\text{TeV}} \right)^{\left[ -(2.2 \pm 0.1) -  ( 1.4 \pm 0.3 ) \log_{10} (\frac{E}{0.2\text{TeV}})  \right]} %\\ & 
%\text{m}^{-2} \text{s}^{-1} \text{TeV}^{-1}, 
%\end{split}
%\end{equation}
%$
%which gives a reduced $\chi^2_\text{dof}$ value of 1.7. 
%Both the observed and the de-absorbed TeV $\gamma$-ray spectra are shown %together with the GeV $\gamma$-ray spectra 
Both the observed and the de-absorbed TeV $\gamma$-ray spectra are shown in Figure~\ref{fig:gammaSED} in the $\nu F \nu$ representation. 

\vspace{-0.3cm}
%%%%%%%%%
\subsection{Fermi-LAT}
\begin{figure*}[t!]
\vspace{-0.85cm}
 \centering 
%\begin{tikzpicture}
%\node (img) { 
%\includegraphics[width=0.8\linewidth]{BLLac_Rband_5rows_shifted_withXRT_timecut_v3_watermarkICRC.pdf}  %};
\includegraphics[width=0.8\linewidth]{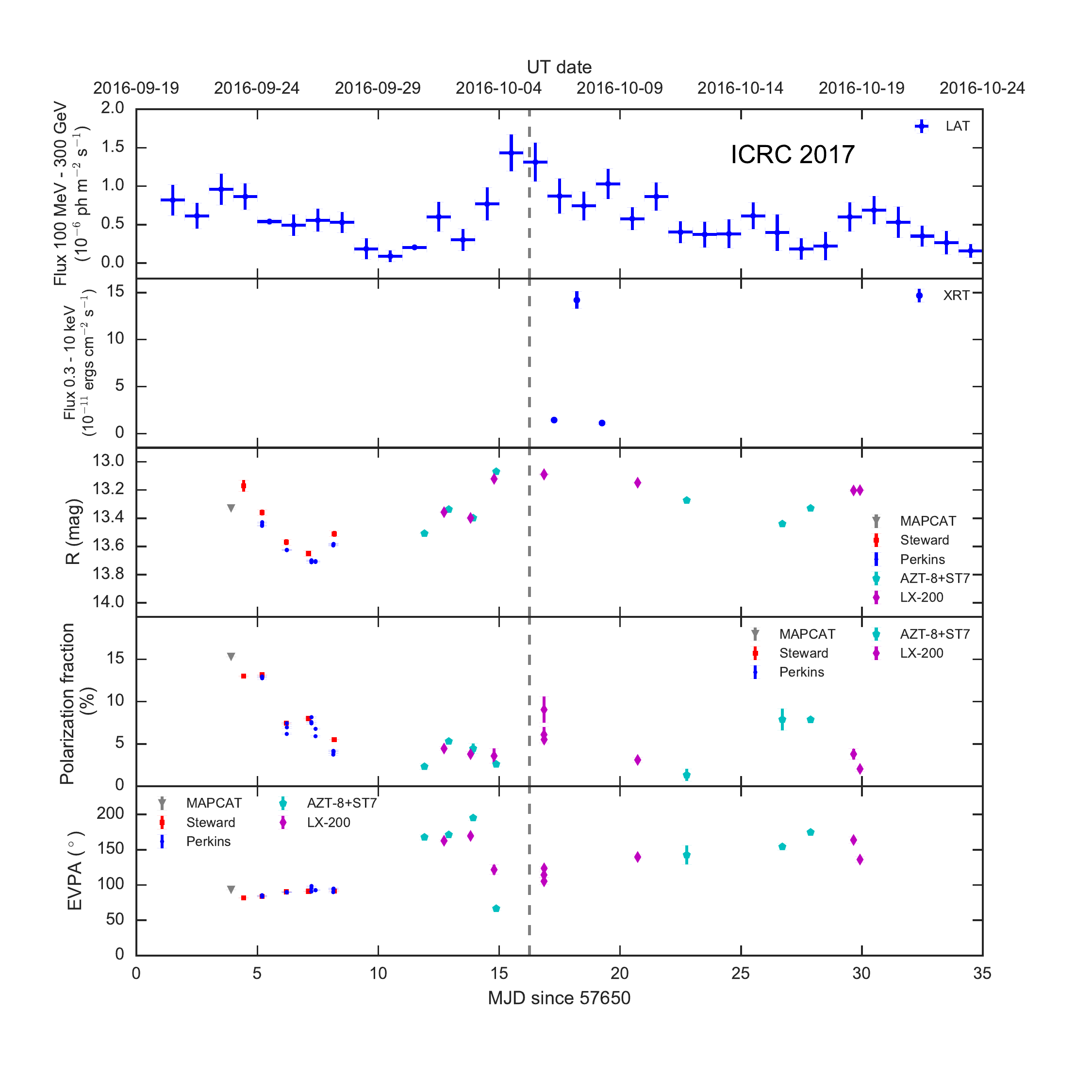}
%\node [text width=3cm, align=center, rotate=45, opacity=0.5, color=orange] {{\Large \bf 2017 ICRC}};
%\end{tikzpicture}
\vspace{-0.85cm}
\caption{The one-month MWL light curves of BL Lac around the time of the VHE flare. The top panel shows the daily-binned GeV $\gamma$-ray light curve measured by {\it Fermi}-LAT. The second panel shows the daily-binned X-ray light curve measured by {\it Swift}-XRT. The lower three panels show the R-band photometric and polarimetric measurements taken by four instruments. The grey dashed line shows the peak time of the TeV flare observed by VERITAS. 
\label{fig:mLC1}}
\vspace{-0.5cm}
\end{figure*}

\vspace{-0.15cm}
%\subsection{High-energy gamma-ray observations with Fermi-LAT}
%%%%%%%%%
The Large Area Telescope (LAT) onboard the Fermi satellite is a pair-conversion $\gamma$-ray telescope sensitive to energies from $\sim$20~MeV to $>$300~GeV \cite{Atwood09}. 
%With the recently developed event-level analysis (Pass 8) by the Fermi-LAT collaboration \cite{Atwood13} 
%
An unbinned likelihood analysis was performed with the LAT ScienceTools \texttt{v10r0p5} and Pass-8 \texttt{P8R2\_SOURCE\_V6\_v06} instrument response functions, %\cite{Atwood13}, 
between 100 MeV and 300 GeV within 10$^\circ$ from the position of BL Lac. 
%For the short durations of interest to the TeV flare, a simple model containing BL Lac, another point source 3FGL~J2151.6+4154 at $\sim2^\circ$ away from BL Lac, and the contributions from the galactic (\texttt{gll\_iem\_v06}) and isotropic (\texttt{iso\_P8R2\_SOURCE\_V6\_v06}) diffuse emission was used. We checked in the residual test-statistics map that no significant excess was left unconsidered by the model. A power-law spectral type was chosen to model BL Lac instead of the log-parabola model in the \texttt{3FGL} catalog for the short durations. 
%
%For the light curve shown in the first panel of Figure~\ref{fig:mLC1}, unbinned likelihood analysis was performed on each one-day interval, leaving the normalizations and power-law indices of BL Lac and 3FGL~J2151.6+4154 free, as well as the normalization of the diffuse sources. 
As shown in the light curve in the first panel of Figure~\ref{fig:mLC1}, BL Lac was in an elevated GeV $\gamma$-ray state when the TeV flare occurred. Its GeV flux varied by a factor of $\sim$2 on a $\sim$1 day timescale. 
%
\iffalse
\begin{figure}[ht!]
 \centering 
 \leavevmode 
 %\includegraphics[width=0.6\textwidth]{BLLac_gamma_SED_2016flare_logpara_deabs.pdf}
  \includegraphics[width=0.6\textwidth]{BLLac_Xray_gamma_SED_2016flare_logpara_deabs.pdf}
 \caption{The $\gamma$-ray and X-ray SEDs measured by {\it Fermi}-LAT, VERITAS, and {\it Swift}-XRT. The {\it Fermi}-LAT SEDs strictly simultaneous with VERITAS observations on 2016 Oct 5 and from the three days around it are shown in blue and grey, respectively. %The observed and de-absorbed VERITAS SED averaged over all observations on the flaring night is shown in red and green, respectively. 
The shaded regions represent 1-$\sigma$ confidence intervals derived from the best-fit spectral model. 
%The X-ray SEDs de-absorbed with the best-fit neutral hydrogen column density values on 2016 Oct 6, 7, and 8 are shown in magenta, cyan, and light blue, respectively. 
 %power-law model from the LAT unbinned likelihood analysis. 
 \label{fig:gammaSED}}
\end{figure}
%
\fi

The GeV $\gamma$-ray SEDs %measured by {\it Fermi}-LAT and VERITAS 
strictly simultaneous with the TeV flare, as well as over a three-day interval around the time of the flare, are shown in Figure~\ref{fig:gammaSED}, 
with the best-fit power-law indices %of BL Lac for the strictly simultaneous and the three-day LAT data are 
of $1.83\pm0.21$ and $1.85\pm0.07$, respectively. 
%Both the GeV and TeV $\gamma$-ray spectral indices of this flare in 2016 are comparable to those of the flare in 2011. 
Both the GeV and TeV $\gamma$-ray spectral indices obtained for this 2016 flare are comparable to those estimated for the 2011 flare.
%
%%%%%%%%%
%
\vspace{-0.3cm}
\subsection{Swift XRT}
\label{subsec:XRT}
%%%%%%%%%
\vspace{-0.15cm}
The X-Ray Telescope (XRT) onboard the Swift satellite is a grazing-incidence focusing X-ray telescope sensitive to photons in the 0.2--10~keV energy range \cite{Burrows05}. % \cite{Gehrels04, Burrows05}. 
Swift followed up on BL Lac on 2016 Oct 6, 7, and 8, and no other observations were made in the one-month period around the time of the VHE flare. 
The XRT data, taken in the photon counting (PC) mode, were analyzed using the \texttt{HEAsoft} package (v6.19). % and \texttt{XSpec}. 
%The event files are calibrated and cleaned using the calibration files from 2011 September 5. The data were taken in the photon counting (PC) mode, and were selected from grades 0 to 12 over the energy range 0.3-10~keV. 
%The data were first processed using \texttt{xrtpipeline} (v0.13.2). 
%Pile-up correction was checked for all observations by fitting a King function to the point spread functions (PSFs) at $>$15", and excluding the central pixels where the data fall below the model, indicating pile-up. 
%
%For the observations on 2016 Oct 6, the King function agrees with the data even on the brightest pixels. Therefore a circular source region of a radius of 20 pixels centered on BL Lac was used. 
Pile-up correction was necessary for data taken on 2016 Oct 7 and 8, and annular source regions with inner radii of 4 and 2 pixels, and outer radius of 20 pixels were used. 
%For all three observations, an annular background region with an inner and outer radii of 70 and 120 pixels, respectively, were used. 
%Note that source regions excluding the central 2 and 4 pixels were also tested for the observations on 2016 Oct 6, and consistent results were obtained. Therefore we are confident that no bias was introduced by the different exclusion regions for pile-up correction. 
The observations on Oct 7 consisted of two intervals of duration 486 seconds and 1422 seconds, the latter of which was discarded as a sustained dark stripe near the position of BL Lac contaminates the XRT image. %separated by roughly one satellite orbital period ($\sim$90 minutes). A sustained dark stripe in the XRT image near the position of BL Lac contaminates the second interval. Therefore only the data during the first interval was used. %The image and spectrum of every $\sim$3 minutes during this relatively short interval were checked for data quality, and no anomaly was found. 
%
\iffalse
%\begin{deluxetable}{ccccc}[h]
%\tablecolumns{5}
\begin{table}[h]
\centering
\begin{tabular}{ccccc} 
%\tablewidth{700pt}
%\tabletypesize{\scriptsize}
\hline
	Date 	&	$\alpha$		& K 						& N$_H$ 			& $\chi^{2}_\text{dof}$ \\
 			&				&  $10^{-2}$ keV$^{-1}$cm$^{-2}$s$^{-1}$ & $10^{21}$ cm$^{-2}$ &  \\
\hline
       Oct 5		& $ 2.5 \pm 0.1 $ 	&  $ 0.62^{+0.07}_{-0.06} $ 	& $ 2.7\pm 0.3 $	& 0.83 	\\
       %Oct 6 		& $ 2.0 \pm 0.2 $	&  $ 1.6^{+0.3}_{-0.2} $ 	& $ 2.8^{+0.6}_{-0.5} $	& 0.97 	\\
       Oct 6 		& $ 2.1 \pm 0.1 $	&  $ 4.6^{+0.6}_{-0.5} $ 	& $ 3.1^{+0.5}_{-0.4} $	& 1.07 	\\
       Oct 7 		& $ 2.3 \pm 0.1 $	&  $ 0.43^{+0.06}_{-0.05} $	& $ 3.0^{+0.5}_{-0.4} $	& 0.54 	\\
%\enddata
%\end{deluxetable}
\hline
\end{tabular}
\caption{ Swift XRT spectral fit results using the absorbed-power-law model described in Equation~\ref{eqXspec}. The errors denote 68\% confidence intervals. }
\label{tab:Xspec}
\end{table}
%
%\iffalse
\begin{figure}[ht!]
 \centering 
 \leavevmode 
 \includegraphics[width=0.6\linewidth]{BLLac_XRT_3day_SED_Oct7_timecut.pdf}
 \caption{The X-ray SEDs measured by {\it Swift}-XRT on 2016 Oct 6, 7, and 8. 
 \label{fig:XSED}}
\end{figure}
\fi

%Ancillary response files were generated using the \texttt{xrtmkarf} task with the response matrix file \texttt{swxpc0to12s6\_20130101v014.rmf}. 
The X-ray spectrum was fitted with an absorbed power law model (\texttt{po*wabs}), and then de-absorbed with the best-fit neutral hydrogen column density values. 
%\begin{equation}
%\label{eqXspec}
%\frac{dN}{dE} = e^{-N_H\sigma(E)} K \left( \frac{E}{1\;\text{keV}} \right)^{-\alpha}, 
%\end{equation}
%where $N_H$ is the column density of neutral hydrogen, $\sigma(E)$ is the photo-electric cross-section, $K$ and $\alpha$ are the normalization and the index of the power-law component, respectively. 
%The best-fit value of parameters are shown in Table~\ref{tab:Xspec}. %Note that the best-fit values of $N_H$ are in agreement with the results reported in \cite{Arlen13} and \cite{Ravasio03}, but were larger than the value $N_H = 0.18  \times10^{22} \;\text{cm}^{-2}$ from the Leiden/Argentine/Bonn (LAB) survey of galactic HI \cite{Kalberla05}. 
The de-absorbed X-ray SEDs of BL Lac on 2016 Oct 6, 7, and 8 are shown in Figure~\ref{fig:gammaSED}. 
The X-ray emission from the source was the strongest and hardest on 2016 Oct 7 (two days after the TeV $\gamma$-ray flare) compared to the day before and after. 
The energy flux values on the three nights were $(1.4\pm 0.1)$, $(14.2\pm 0.9)$, and $(1.1\pm 0.1)$ $\times10^{-11}$ ergs cm$^{-2}$ s$^{-1}$, as shown in the second panel of Figure~\ref{fig:mLC1}. 

\vspace{-0.4cm}
%%%%%%%%%
\subsection{Optical facilities}
%\subsection{R-band polarization}
%%%%%%%%%
%
\vspace{-0.2cm}
BL Lac was monitored in R band at a high cadence by a number of optical facilities, including the Steward Observatory\cite{Smith09}\footnote{\url{http://james.as.arizona.edu/~psmith/Fermi}}, the AZT-8 reflector of the Crimean Astrophysical Observatory, the Perkins telescope\footnote{\url{http://lowell.edu/research/research-facilities/1-8-meter-perkins/}}, the LX-200 telescope in St. Petersburg, Russia, and the Calar Alto Telescope under the MAPCAT\footnote{\url{www.iaa.es/~iagudo/research/MAPCAT/MAPCAT.html}} program. 

The R-band flux and polarization measurements around the time of the VHE $\gamma$-ray flare are shown in Figure~\ref{fig:mLC1}. 
The lower three panels show the R-band magnitude, polarization fraction, and electric vector position angle (EVPA), respectively. 
%The fifth panel shows the difference between each EVPA and the previous one. 
A 180$^\circ$ shift is added to all the EVPA measurements after MJD 57662, so that the difference between adjacent EVPA measurements %between MJD 57662 and 57658 is reduced to $\sim$80$^\circ$ from $\sim$100$^\circ$ before the shift was applied. 
is minimized. 
%The measurements from all instruments agree with one another.  
%
%The grey dashed line shows the peak time of the TeV flare observed by VERITAS. 
An increase in the R-band flux, accompanied by a decrease of the optical polarization fraction and a deviation in the EVPA by $\sim$90$^\circ$ were observed from the source a few days before the VHE flare. Such behaviour is consistent with the emergence of a radio knot observed by VLBA described below. 
\begin{figure*}[t]
%\centering
\vspace{-0.85cm}
\hspace{-0.8cm}
%\begin{tikzpicture}
%\node (img) {   \includegraphics[width=1.1\textwidth]{BLLAC_vlba_fall2016.eps} };
\includegraphics[width=1.1\textwidth]{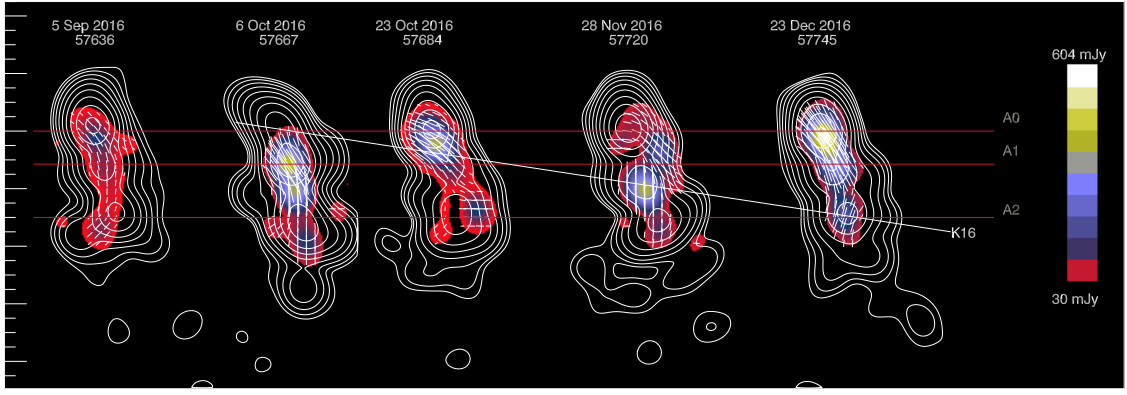}
%\node [text width=3cm, align=center, rotate=45, opacity=0.5, color=orange] {{\Large \bf 2017 ICRC}};
%\end{tikzpicture}
%\fig{BLLAC_vlba_fall2016.eps}{0.58\textwidth}{}
\vspace{-0.85cm}
\caption{Total (contours) and linearly polarized (color code) intensity images of BL Lac from VLBA observations at 43 GHz. The images are convolved with a circular Gaussian restoring beam of FWHM = 0.1 mas. Tick marks are separated by 0.05 mas. %The contours decrease in multiples of 1/$\sqrt{2}$ from the peak of 1.15 Jy/beam. 
The contours increase in multiples of 2 starting from 0.5\% of the peak of 1.15 Jy/beam.
Red horizontal lines indicate the mean locations of the three quasi-stationary components, while the white line shows the motion of moving knot $K16$.
\label{fig:VLBA}}
\vspace{-0.4cm}
\end{figure*}

\vspace{-0.25cm}
%%%%%%%%%
\subsection{VLBA}
%%%%%%%%%
\label{sec:VLBA}
%BL Lac was regularly observed with VLBA at 43~GHz and 86~GHz as part of the monitoring program of gamma-ray bright blazars at Boston University \footnote{\url{http://www.bu.edu/blazars/VLBAproject.html}}. 
%A series of 43-GHz VLBA images of BL Lac were taken on 2016 Jul 31, Sep 5, Oct 6, Oct 23, and Nov XX. 
%86~GHz ~ 3.5 mm; 43~GHz ~ 7 mm
%
%The data were correlated at the National Radio Astronomy Observatory in Socorro, NM, and analyzed at Boston University following the procedures described by \cite{Jorstad05}. 
%No significant change in the structure of the inner jet was observed from Jul 31 to Sep 5, 
%while changes between Sep 5 and Oct 6, as well as between Oct 6 and Oct 23, were observed. 
%
%...(more after VLBA data are delivered, core shift as a function of frequency + RM)
%
BL~Lac was observed throughout the period of interest at 43 GHz with the VLBA under the VLBA-BU-BLAZAR monitoring program \cite{Jorstad16} and at 15.4 GHz under the Monitoring Of Jets in Active galactic nuclei with VLBA Experiments (MOJAVE) program \cite{Lister09}, %. The 43-GHz and 15.4-GHz VLBA 
the data calibration and imaging procedures of which were identical to those described by \cite{Jorstad05} and \cite{Lister09}, respectively. 

%BL Lac is also in the sample of the Monitoring Of Jets in Active galactic nuclei with VLBA Experiments (MOJAVE) program. For this work, we only used results from polarization measurements at 15.4 GHz. The data reduction procedures are described by \cite{Lister09}. Briefly, the flux density of the core component is derived from a Gaussian model fit to the interferometric visibility data. Polarization properties of the core are then derived by taking the mean Stokes Q and U flux densities of the nine contiguous pixels that are centered at the Gaussian peak pixel position of the core fit. The results include fractional linear polarization, electric vector position angle (note the 180~degree degeneracy), and polarized flux densities. The flux density has an uncertainty of $\sim 5\%$, and the position angle of polarization has an uncertainty of $\sim 3$~degrees.

Figure~\ref{fig:VLBA} presents images of the parsec-scale jet of BL~Lac at five epochs from 2016 September 5 to December 23. %The second epoch, 2016 October 6, took place only one day after the VHE flare. 
%The images are convolved with a circular Gaussian restoring beam with a FWHM of 0.1 milliarcsec (mas), which is similar to the angular resolution of the longest baselines along the (southern) direction of the jet. 
%We note that the October 6 observation was plagued by equipment failure at the Mauna Kea and Hancock antennas, at the extremities of the array. However, this degraded the north-south angular resolution by only 14\%. 
%The corresponding 
The linear resolution at the redshift of BL~Lac is 0.13 pc (corresponding to 0.1~mas) in projection on the sky and $1.8^{+0.8}_{-0.4}$ pc for a viewing angle of $4.2^\circ\pm1.3^\circ$ between the jet axis and line of sight \cite{Jorstad17}.

As was the case in earlier observations \cite{Jorstad05, Arlen13, Wehrle16}, the main structure of the compact jet consists of three quasi-stationary brightness peaks, designated as (from north to south) $A0$ (used as the positional reference point), $A1$ 0.12 mas to the south, and $A2$ 0.30 mas to the south. The locations of $A1$ and $A2$ appear to fluctuate as moving emission features with superluminal apparent velocities pass through the region. This combination of moving and stationary emission components complicates interpretation of the changing total and polarized intensity structure. %, making our interpretation not unique. 
Therefore, the interpretation that we offer to explain the variations seen the images is not unique. 
%We ignore the effects of Faraday rotation on the polarization EVPA, which \cite{Jorstad07} estimated to be low ($-16^\circ$) at 43 GHz.

A knot of emission with enhanced polarization, designated as $K16$, appears to propagate down the jet. It moves by 0.23 mas between October 23, when its centroid is $\sim 0.05$ mas south of $A0$, and December 23, when it is 0.28 mas from $A0$. This corresponds to an apparent speed of $6c$, within the range typically observed in BL~Lac \cite{Arlen13, Marscher08, Jorstad05, Jorstad17, Wehrle16}. %\cite{Jorstad05, Marscher08, Arlen13, Wehrle16, Jorstad17}. 
Extrapolation back to October 6 places the knot 0.01 mas north of the centroid of $A0$, within the $A0$ emission region given its angular size of $0.03\pm0.02$ mas \cite{Jorstad17}. This interpretation implies that the VHE flare occurred as the knot crossed the ``core,'' which has been interpreted as a standing shock located $\sim1$ pc from the black hole \cite{Marscher08}.
%
%\begin{figure}[ht!]
\begin{figure*}[t]
\vspace{-0.85cm}
%\hspace{-1cm}
%\gridline{\fig{2200+420_15GHz_1.pdf}{0.51\textwidth}{}}
%\gridline{\fig{2200+420_15GHz_2.pdf}{0.51\textwidth}{}}
%\begin{tikzpicture}
%\node (img) {   \includegraphics[width=0.47\textwidth]{2200+420_15GHz_1_no_cb.pdf}
% \includegraphics[width=0.52\textwidth]{{2200+420.ip2_v2}.pdf} };
% \vspace*{-0.35cm}
%\node [text width=3cm, align=center, rotate=45, opacity=0.5, color=orange] {{\Large \bf 2017 ICRC}};
% \vspace*{-0.35cm}
%\end{tikzpicture}
 \includegraphics[width=0.47\textwidth]{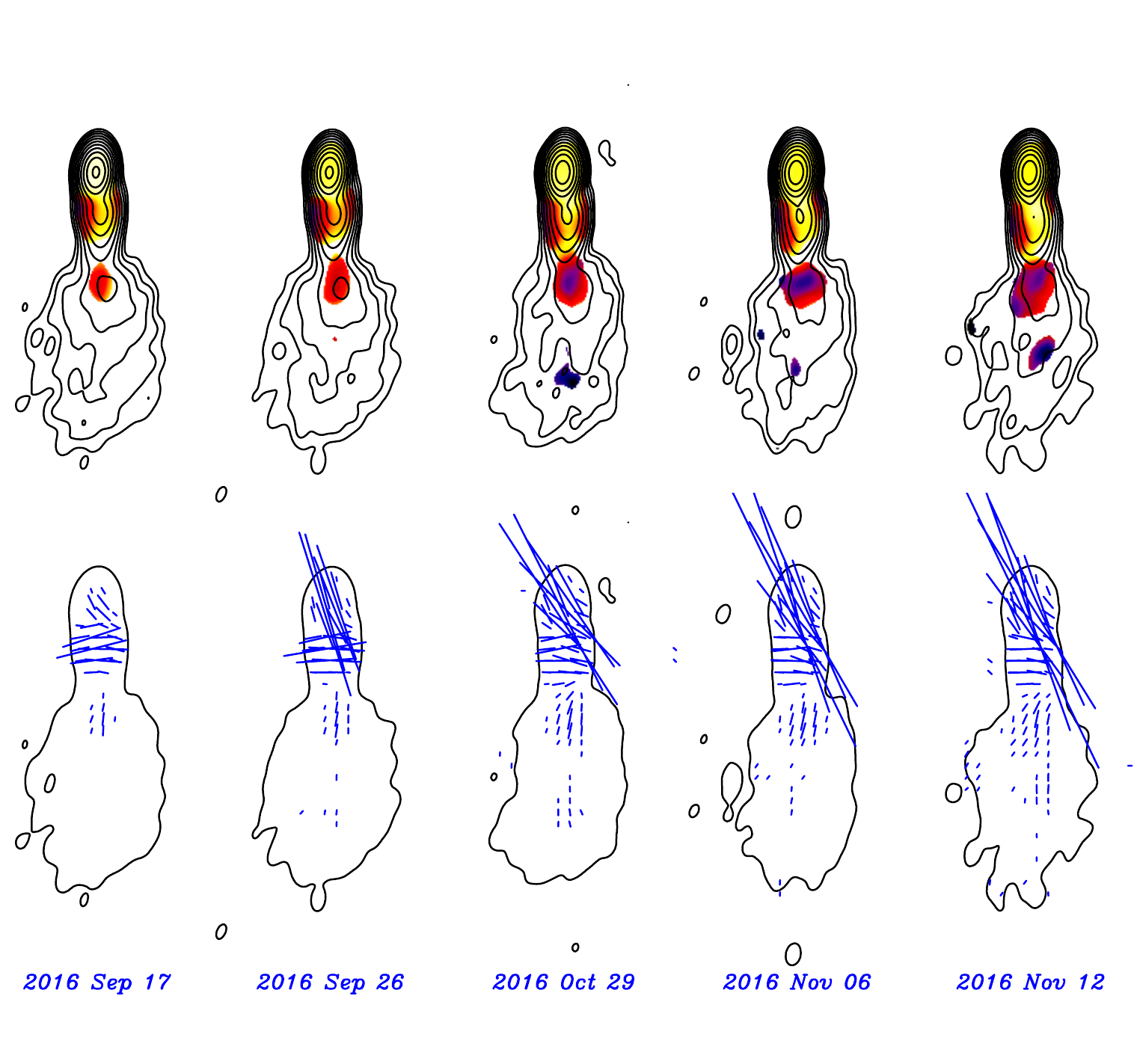}
 \includegraphics[width=0.52\textwidth]{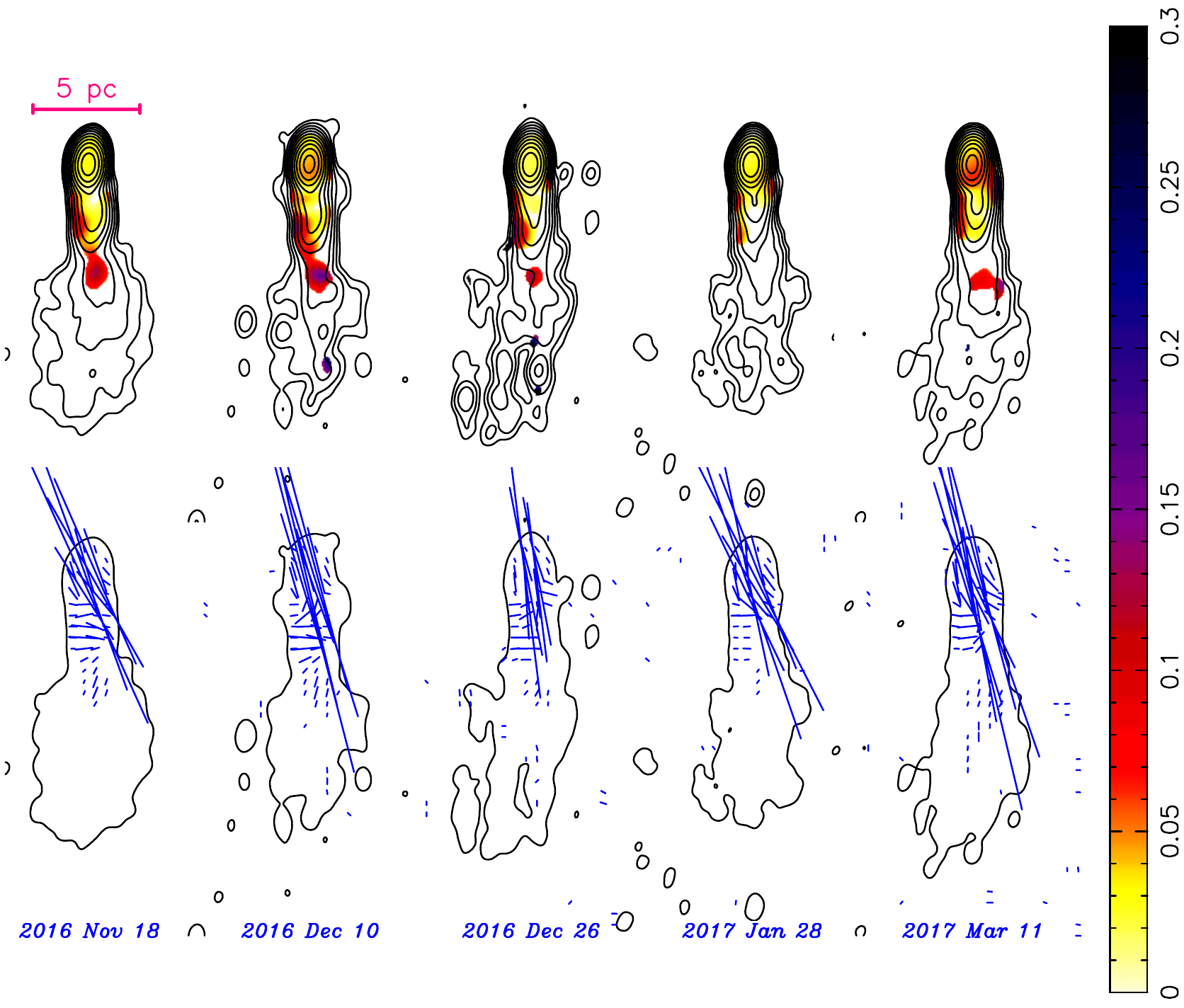}
% \vspace*{-0.35cm}
\caption{Images of BL Lac from VLBA observations at 15.4 GHz at nine epochs in 2016. A Gaussian restoring beam with dimensions 0.883 $\times$ 0.56 mas and a position angle $-8.2$ degrees was used. 
The contours show the total intensity, with a base contour of 1.1 mJy/beam and successive contours in the top row incrementing by factors of two. The colors in the top rows show the fractional polarization, and the blue line segments in the bottom rows show the EVPA. 
The length of the EVPA line segments corresponds to polarized intensity, the lowest of which shown is 0.5 mJy/beam. %plotted every 8th pixel, The pixel scale is 0.05 mas/pixel
The typical total and polarized intensity image rms are 0.09 mJy/beam and 0.1 mJy/beam, respectively. 
%The angular scale of the image is 1.29 pc/mas. The polarization color scale in each image is shown on the side. 
\label{fig:VLBA15}}
\vspace{-0.4cm}
\end{figure*}
%\end{figure}

The VLBA images at 15.4~GHz reveal the evolution of jet structures further away from the central source on a larger spatial scale compared to the 43 GHz images, as a result of a steep spectrum of optically thin synchrotron emission of outer jet regions. %optical depth and angular resolution. %, respectively. 
%Therefore a delay is expected between the measurements at these two frequencies. 
Figure~\ref{fig:VLBA15} shows that the fractional polarization of the stationary core of BL Lac at 15.4 GHz suddenly dropped on 2016 Dec 26, and gradually increased in the two epochs afterward. A small region with enhanced polarized intensity and distinct EVPA south of the core appeared on Dec 26, which may correspond to the knot $K16$ observed at 43 GHz. However, we note that there is no indication of the same region in the total intensity map at 15.4 GHz. 
%This is consistent with past observations of BL Lac with VLBA at different frequencies \cite{Bach06}, where new components of the jet were detected to fade as they separated from the core, then disappear between about 0.7 and 1 mas, and reappear at $\sim$1~mas. 

%%%%%%%%%%%%%%%%%%
%%%%%%%%%%%%%%%%%%
%%%% %     Discussion     %%%%
%%%%%%%%%%%%%%%%%%
%%%%%%%%%%%%%%%%%%
\vspace{-0.6cm}
\section{Discussion}
\label{sec:discuss}
\vspace{-0.3cm}
For the second time, a fast TeV $\gamma$-ray flare from BL Lac was observed coincidentally with the appearance of a candidate superluminal radio knot. 
This suggests a possible association between the fast VHE $\gamma$-ray flare and the appearance of the superluminal radio knot for the source, similar to that reported by \cite{Arlen13}. 
%While it is difficult to quantify the significance of such an association in a single occurrence of the two events, the chance of them being random coincidence reduces as more events are observed. 
% with 
%In contrast, the VERITAS measurements during the previous VHE fast flare from the object did not provide information about the rising phase. 
%While no information from the rising phase of the first VHE flare was obtained \cite{Arlen13}, the VERITAS measurements during the 2016 flare in this work cover both the rise and decay phases of the flare. 
%Duty cycle of the jet shooting blobs? 
%Are these two flares the same or different? GeV, X-ray?

%\vspace*{-0.05cm}
The fastest timescale of a flare (in this case the decay time) can be used to put an upper limit on the size of the emitting region, as %$R\le\text{c}t_\text{decay} \delta /(1+z)$, 
%\begin{equation}
%\label{eqR}
%R\le \frac{\text{c}t_\text{decay} \delta} {1+z}, 
%\end{equation}
$R\le \text{c}t_\text{decay} \delta/(1+z)$, 
where $c$ is the speed of light, $z$ is the redshift of the source, and $\delta$ is the Doppler factor of the jet \footnote{$\delta = [\Gamma (1-\beta cos \theta)]^{-1}$, where
$\Gamma$ is the bulk Lorenz factor of the jet and $\theta$ is the viewing angle.}. 
%
%The mass of the central black hole ($M_\text{BH}$) of BL Lac was estimated to be $\sim3.8\times 10^8 M_\odot$ by \cite{Wu09} using the R-band absolute magnitude and the empirical correlation between black hole mass and bulge luminosity of the host galaxy \cite{McLure02}, %\cite{Xie04}, \cite{Gupta12} 
Taking the values of the mass of the central black hole $M_\text{BH}\sim3.8\times 10^8 M_\odot$ \cite{Wu09}, %the Schwarzschild radius $R_s$ of the central black hole of BL Lac is $\sim1.17\times 10^{12}$ m (or $\sim 3.8\times 10^{-5}$ pc). 
%3.09e-08 arcsec
%It is worth noting that the mass measurement of a black hole is a challenging task, and a range of $M_\text{BH}$ was reported as $(0.18 - 5.01)\times 10^8 M_\odot$ \cite[see][and references there in]{Gupta12}. 
%
the Doppler factor $\delta \sim$24 \cite{Hervet16}, %from the propagation of a possible perturbation in the radio jet observed with VLBA at 15 GHz~\cite{Lister13}, assuming a viewing angle of $2.2^\circ$ based on radio apparent velocity measurement. Taking 
and the best-fit $t_\text{decay}=36^{+8}_{-7}$~minutes, %(see Subsection~\ref{subsubsec:VHELC}), 
the upper limit of the size of the emitting region is estimated %using Equation~\ref{eqR} 
as $R \lesssim11.9 R_s$, where $R_s$ is the Schwarzschild radius. %$R \le 11.9^{+2.6}_{-2.3} R_s$. 
%Similarly, the much shorter best-fit decay timescale in model II leads to a stronger constraint on the size of the emitting region: $R \le 0.9^{+2.2}_{-0.3} R_s$. 
% model I MCMC $ 11.9^{+2.6}_{-2.3} $ 
% t_decay = 1740 s (29 min), delta=23 => R<= 1.12e13m ~ 3.6e-4 pc ~ 9.6 R_Schwarzschild
% t_decay = 2160 s (36 min), delta=23 => R<= 1.39e+13m ~ 4.5e-4 pc ~ 11.9 R_Schwarzschild
% t_decay = 2640 s (44 min), delta=23 => R<= 1.7e13m ~ 5.5e-4 pc ~ 14.5 R_Schwarzschild
% model II MCMC $ 0.9^{+2.2}_{-0.3} $
% t_decay = 108 s (1.8 min), delta=23 => R<= 7e11m ~ 2.3e-5 pc ~ 0.6 R_Schwarzschild
% t_decay = 156 s (2.6 min), delta=23 => R<= 1.0e+12m ~ 3.3e-5 pc ~ 0.86 R_Schwarzschild
% t_decay = 558 s (9.3 min), delta=23 => R<= 3.6e12m ~ 11.7e-5 pc ~ 3.1 R_Schwarzschild
%Both model I and model II are justified, model I is motivated by the Occam's razor, and model II by multi-zone scenario. 
%
%
%calc_dL 0.069
%Luminosity distance: ouput values = 
%age at z, 	distance in Mpc, 	kpc/arcsec, 	apparent to abs mag conversion
%11.71 		271.47 			1.23 			37.31
% 1.23 pc/mas
%Note that model II is statistically preferred taking into account its better agreement with data, as well as the extra parameter (see Appendix). 

%Slow rise and fast decay, what does it tell us? Sign of a brutal stop of the injection? 
%An asymmetric profile with a rather uncommon faster decay of the VHE $\gamma$-ray flux were observed in the flare, which 
The decay time of the observed VHE $\gamma$-ray flare was faster than the rise time. This is uncommon and 
may be caused by an abrupt stop of the high-energy particle injection \cite[see e.g.][]{Katarzynski03}. 
%In this scenario, the flaring activity was caused by fresh injection of high-energy particles into the emitting region instead of in situ acceleration of the particles. However, minimal variability in radio band is expected. 
%The asymmetric flare is in contrast with the more often observed flaring profile, a fast rise followed by a slow decay, which can be the manifestation of in situ acceleration and/or a longer cooling time (than acceleration time) associated with a soft particle energy distribution \cite[analogous to solar flares; see e.g.][]{Harra16}. 
%
The enhancement in the GeV $\gamma$-ray and X-ray flux of BL Lac contemporaneous with the TeV flare provides evidence for strong activity of the relativistic particles in the jet, but not enough information for any abrupt stop. 

%Flare is consistent with magnetic reconnection. 
%Destructuration of the jet: a magnetic reconnection event can break the MHD jet structure 
%A superluminal radio knot $K16$ was observed by a series of VLBA exposure on BL Lac at 43 GHz. 
%The extrapolation of the position of the candidate superluminal radio knot $K16$ observed by VLBA at 43 GHz implies that the VHE $\gamma$-ray flare happened as the knot $K16$ crossed the quasi-stationary radio core. 
%This suggests a possible association between the fast VHE $\gamma$-ray flare and the appearance of the superluminal radio knot for the source, similar to that reported by \cite{Arlen13}. 
%This is the second detection of a fast TeV $\gamma$-ray flare and the appearance of a superluminal radio knot coincidental in time from BL Lac, making it less likely to be a random coincidence. 

In the model proposed by \cite{Marscher14}, the radio core is a structure containing a conical shock, perhaps with a Mach disk at its apex, downstream of the base of the jet. % with a transverse orientation with respect to the jet axis. 
Turbulent cells of plasma first pass through the conical shock, where electrons are accelerated. % to relativistic speed. %and produce a $\gamma$-ray flare through inverse-Compton scattering. 
The plasma then passes the Mach disk, where a fast $\gamma$-ray flare can happen via inverse-Compton scattering. 
%
%When turbulent cells of plasma pass through the conical shock, electrons can be accelerated to relativistic speed. %and produce a $\gamma$-ray flare through inverse-Compton scattering. 
%A fast $\gamma$-ray flare can happen as the relativistic plasma passes the Mach disk at the end of the conical shock via inverse-Compton scattering. 
%The slow but highly compressed plasma in the Mach disk provides a highly variable local source of seed photons for inverse Compton scattering by electrons in the faster plasma that passes across the conical shock. If a region of especially high density of relativistic electrons passes through the core, it can cause a sharp flare at $\gamma$-ray energies and 
After the Mach disk, a conical rarefaction can cause the plasma flow to expand and accelerate and appear as a superluminal radio knot. 
%
%This model is able to produce sharp flares, as well as variations in polarization angle and fraction, consistent with the observations. 
Polarization changes including a drop in the polarization fraction (due to the new magnetic field of the passing plasma cancelling that of the stationary core) and a swing in the polarization angle (as the passing plasma becomes brighter and dominant over the stationary core) are predicted in a similar model \cite{Zhang14}. %$\gamma$-ray flares in similar models \cite[e.g.][]{Marscher14, Zhang14}. 
This is consistent with the optical polarization measurements from shortly before the VHE $\gamma$-ray flare, % (see Figure~\ref{fig:mLC1}), 
as well as the VLBA images at 43 and 15.4~GHz afterward.  %(see Figure~\ref{fig:VLBA15}). %The changing superposition of the magnetic fields as the passing knot ($K16$) moves along the quasi-stationary knots ($A0$, $A1$, and $A2$) may also explain the change in the positions of $A0$, $A1$, and $A2$ between epochs (see Figure~\ref{fig:VLBA}). 
%Although we note that in this model the flares are caused by continuous noise processes of turbulent instead of singular events such as explosive injection of relativistic particles at the base of the jet, in contrast with the model we invoked above \cite{Katarzynski03} to explain the sharp fall of the flux. 
%Such an association was supported by the optical and radio polarization evolution around the time of the flare. 

%
An alternative explanation of both the VHE $\gamma$-ray flare and the superluminal radio knot of BL Lac %the break out of a stationary knot at a recollimation shock 
is the breakout of a recollimation shock zone %(observed as a stationary knot before the breakout) 
\cite{Hervet16}. %, is also possible. 
In this model, %one or more recollimation shocks, of similar nature as those in \cite{Marscher14}, can form upstream the jet where the magnetic energy density is high and appear as stationary radio knots; while at further downstream of the jet, particle kinetic energy becomes dominant, magnetic field becomes unstable, and 
a stationary knot can be carried away by the relativistic underlying flow and become a superluminal knot. 
%In the case of a compact region with large kinetic energy passing the recollimation shock zone, a sequence of events could occur and lead to a multi-component flare with one component behaving like a constant baseline on short timescales of hours. First, an increase in the non-thermal emission of the shock region is expected, which can lead to a flux increase on the timescale corresponding to the size of the entire shock region (as the baseline component). 
%As the kinetic power of the jet increases at the shock zone, the relative strength of the magnetic field decreases until the magnetic field structure can no longer be supported, at which point 
During this process, a magnetic reconnection event can occur, leading to the observed fast flare. 
%Finally, the shock zone is dragged away by the flow and enters an adiabatic expansion and cooling phase, leading to a decrease in flux and return to the low state of the source. 
In the case of the 2016 flare of BL Lac, there is no evidence for the disruption or breakout of a stationary knot. Although it is possible that the recollimation zone reformed quickly between VLBA epochs and therefore wasn't sampled by the observations. 

%The current data do not allow a unique interpretation of the association between them. 

%The observed fast flare can also be explained by magnetic reconnection \cite{Giannios09}. In this model, material with high magnetization can occasionally advect into small regions within the bulk of a magnetically dominated jet, and dissipate energy through magnetic reconnection, forming a ``jet-in-a-jet'' that is relativistic in the jet frame. Such a reconnection region is observed in the lab frame at a much higher Lorentz factor compared to the bulk of the jet, and is able to produce a fast TeV gamma-ray flare through inverse-Compton process. Note that an simultaneous X-ray flare is naturally predicted in this model. 

%...(more after VLBA data are delivered)
%The emergence of a new component observed ??

%Where was the flare produced, we don't know. 
%Radio knows where. Maybe TeV is related to radio: the same blob manifests activities observed in both band. 

%Polarization in radio and optical support this relation. 

%
\vspace{-0.5cm}
\acknowledgments
\vspace{-0.2cm}
%{\small
VERITAS is supported by grants from the U.S. Department of Energy Office of Science, the U.S. National Science Foundation and the Smithsonian Institution, and by NSERC in Canada. We acknowledge the excellent work of the technical support staff at the Fred Lawrence Whipple Observatory and at the collaborating institutions in the construction and operation of the instrument.
The VERITAS Collaboration is grateful to Trevor Weekes for his seminal contributions and leadership in the field of VHE $\gamma$-ray astrophysics, which made this study possible.

The research at Boston University was supported in part by NASA Fermi Guest Investigator Program grant NNX14AQ58G. The VLBA is an instrument of the Long Baseline Observatory (LBO). The LBO is a facility of the National Science Foundation operated under cooperative agreement by Associated Universities, Inc. 
The MOJAVE program is supported under NASA-Fermi grant NNX15AU76G. 
Data from the Steward Observatory spectropolarimetric monitoring project were used. This program is supported by Fermi Guest Investigator grants NNX15AU81G.
%
%IA acknowledges support by a Ram\'on y Cajal grant of the Ministerio de Econom\'ia y Competitividad (MINECO) of Spain. The MAPCAT program was funded in part by MINECO through grants AYA2010-14844, AYA2013-40825-P, and AYA2016-80889-P, and by the Regional Government of Andaluc\'ia through grant P09-FQM-4784. This paper is partly based on observations carried out at the Calar Alto observatory, which is jointly operated by MPIA (Germany) and the IAA-CSIC (Spain).
%
IA acknowledges support by a Ram\'on y Cajal grant of the Ministerio de Econom\'ia y Competitividad (MINECO) of Spain. 
The MAPCAT program was funded in part by MINECO through grants AYA2010-14844, AYA2013-40825-P, and AYA2016-80889-P, and by the Regional Government of Andaluc\'ia through grant P09-FQM-4784. %This paper is partly based on observations carried out at the Calar Alto observatory, which is jointly operated by MPIA (Germany) and the IAA-CSIC (Spain).

%}

\vspace{-3mm}
\vspace{-0.2cm}
\printbibliography
\vspace{-0.2cm}

\end{document}